# Space group symmetries of the phases of $(Pb_{0.94}Sr_{0.06})(Zr_xTi_{1-x})O_3$ across the antiferrodistortive phase transition in the composition range $0.620 \lesssim x \lesssim 0.940$

Ravindra Singh Solanki[1], Anatoliy Senyshyn[2], and Dhananjai Pandey[1]*

[1]*School of Materials Science and Technology, Indian Institute of Technology (Banaras Hindu University), Varanasi-221005, India*

[2]*Forschungsneutronenquelle Heinz Maier-Leibnitz (FRM II), Technische Universität München, Lichtenbergstrasse 1, D-85747 Garching bei München, Germany*

## Abstract

The existing controversies about the space group symmetries of $Pb(Zr_xTi_{1-x})O_3$ (PZT) above and below antiferrodistortive (AFD) phase transition temperature ($T_{AFD}$) in the $Zr^{4+}$-rich ($0.620 \lesssim x \lesssim 0.940$) compositions are addressed using the results of dielectric, synchrotron x-ray powder diffraction (SXRPD) and neutron powder diffraction (NPD) studies. These compositions undergo an AFD phase transition above room temperature due to tilting of oxygen octahedral leading to a superlattice phase of PZT. We have substituted 6% $Sr^{2+}$ at $Pb^{2+}$-site to enhance the tilt angle and thereby the intensity of the superlattice peaks. The real and imaginary parts of complex dielectric permittivity have been used to locate the paraelectric to ferroelectric and ferroelectric to AFD phase transitions. Rietveld analysis of SXRPD and NPD profiles unambiguously reject the rhombohedral phases in R3c and R3m space groups below and above $T_{AFD}$, respectively, with or without a coexisting monoclinic phase in Cm space group, and confirm that the true symmetries are monoclinic in Cc and Cm space groups below and above $T_{AFD}$, respectively. Based on these and previous findings a phase diagram of PSZT for $0.40 \leq x \leq 0.90$ showing stability fields of monoclinic Cc and monoclinic Cm, tetragonal P4mm and cubic $Pm\bar{3}m$ phases has also been presented.

***Corresponding Author's email:** dp.mst1979@gmail.com

## I. Introduction

Structural phase transitions in perovskites are broadly classified as ferrodistortive (FD) and antiferrodistortive (AFD) types depending on whether they are driven by the freezing of a zone centre ($\vec{k}=0$) or a zone boundary ($\vec{k}\neq 0$) phonon mode [1, 2]. The classic examples of FD phase transitions are the ferroelectric phase of $PbTiO_3$, $KNbO_3$, $BaTiO_3$ resulting from the freezing of the $\Gamma_4^-$ (k=0, 0, 0) mode of the paraelectric cubic $Pm\bar{3}m$ phase [3]. While the AFD transitions are ubiquitous in perovskite, the most investigated example is the tilt transition involving anti-phase rotation of oxygen octahedra in the tetragonal phase of $SrTiO_3$ below $T_c$~105K resulting from the freezing of the $R_4^+$ (k=1/2, 1/2, 1/2) mode of the cubic $Pm\bar{3}m$ phase [4, 5]. AFD phase transitions can also result in in-phase rotation of oxygen octahedra driven by $M_3^+$ (k=1/2, 1/2, 0) mode of the cubic phase as has been reported in $KMnF_3$ [6]. There are several examples of perovskites where both FD and AFD phase transitions occur one after the other at two different characteristics temperatures, such as, in $(Sr_{0.96}Ca_{0.04})TiO_3$, where a ferrielectric phase results from successive freezing of one of the components of the $R_4^+$ mode and $\Gamma_4^-$ mode of the cubic paraelectric phase on lowering the temperature below 225 and 34K, respectively [7]. In the solid solutions of the well known multiferroic $BiFeO_3$, experimental observations suggest that both the $R_4^+$ and $\Gamma_4^-$ modes freeze at the same temperature causing a cubic $Pm\bar{3}m$ to rhombohedral R3c phase transition [8].

The paraelectric cubic phase of the technologically important solid solutions of $PbTiO_3$ with $PbZrO_3$, i.e. $Pb(Zr_xTi_{1-x})O_3$ (commonly known as PZT), also undergoes FD (ferroelectric) and AFD structural phase transitions. In addition, PZT shows a composition induced morphotropic phase transition, from tetragonal phase with P4mm space group stable for $0.00 \leq x < 0.515$ to a pseudorhombohedral monoclinic phase with Cm space group stable for $0.530 < x \leq 0.620$ through a composition range $0.520 \lesssim x \lesssim 0.530$ over which a pseudotetragonal monoclinic phase in the Cm space group is stable at room temperature [9, 10].

There are two different classes of technologically useful PZT ceramics. The PZT compositions lying very close to the morphotropic phase boundary (MPB), across which there is a composition induced tetragonal to monoclinic phase transition, show pronounced piezoelectric activity and are therefore widely used for electromechanical sensing and actuating devices [11]. The PZT compositions away from the MPB on the $Zr^{4+}$-rich side of the MPB, on the other hand, show large pyroelectric coefficient useful for infrared detector applications [11]. The ground state of PZT for the entire range of compositions with tetragonal structure, except those very close to the MPB, is decided by the ferrodistortive transition resulting from the freezing of the polar $\Gamma_4^-$ mode. However, the technologically important PZT compositions are not the ones with the tetragonal ground state. The ground state of both the strongly piezoelectric and pyroelectric PZT compositions is decided not by the FD transition but AFD transition resulting from the freezing of the $R_4^+$ mode involving anti-phase rotation of oxygen octahedra.

In the technologically important piezoelectric ceramic compositions near the MPB, the Curie transition from the paraelectric cubic to ferroelectric tetragonal phase resulting from the freezing of one of the components of the $\Gamma_4^-$ mode is followed by another structural phase transition driven by the freezing of the remaining two components of the $\Gamma_4^-$ mode. This structural transition occurs due to the slightly tilted nature of MPB towards the $Zr^{4+}$-rich side and was investigated in detail by Mishra et al. [12, 13] above room temperature and Noheda et al. [14] below the room temperature. Following the pioneering work by Noheda et al. [14], the low temperature ferroelectric phase resulting from the tetragonal ferroelectric phase is now known to be monoclinic in the Cm space group with a pseudotetragonal character. However, as shown by Ragini et al. [15] and Ranjan et al. [16], this Cm phase is not the ground state of PZT in the MPB region since it undergoes an AFD transition to a superlattice phase which obviously constitutes the ground state of PZT for compositions close to the MPB. For the pyroelectric compositions with $0.620 \lesssim x \lesssim 0.940$ the AFD transition occurs below the Curie point ($T_c$) but above the room temperature, as has been known for several decades [11].

Historically, the structure of the ferroelectric phases of PZT above and below AFD phase transition temperature ($T_{AFD}$) in the composition range $0.620 \lesssim x \lesssim 0.940$ has been believed to correspond to rhombohedral phase in the R3m and R3c space groups [11], respectively. However, following the work of Ragini et al. [9] and Singh et al. [10], it is now widely accepted that the true space group symmetry of the ferroelectric R3m phase is monoclinic in Cm space group, even though it show pseudorhombohedral feature, for $0.530<x<0.620$ at room temperature. Pandey et al. [17] have proposed that the correct space group symmetry of the high temperature ferroelectric phase in the composition range $0.620\lesssim x\lesssim 0.940$ below the Curie point is also Cm. It is the monoclinic phase in the Cm space group that undergoes AFD transition to a superlattice phase for compositions close to the MPB, in which the Cm phase has got a pseudotetragonal feature, as also for the $Zr^{4+}$-rich compositions for which the monoclinic phase has got pseudorhombohedral character.

What is the space group symmetry of the superlattice ground state phase of PZT? The space group of the superlattice phase for the MPB compositions was shown to be Cc by Hatch et al. [18] which was confirmed by several other workers both experimentally [19-22] and theoretically [23]. Subsequently, some workers raised doubts about the correctness of the Cc space group for the superlattice phase and as an alternatively proposed R3c space group. Since this rhombohedral phase in the R3c space group could not explain the splitting of the $(200)_{pc}$ (pc=pseudocubic) peak, they postulated a coexisting monoclinic phase in the Cm space group. The controversy between the Cc and R3c+Cm space group models has been recently resolved unambiguously in favour of the Cc space group using 6% $Sr^{2+}$-substituted [22] and in pure PZT samples [24].

Pandey et al. [17] have proposed that the space group of the superlattice phase below the AFD transition in the composition range $0.620\lesssim x\lesssim 0.940$ should also be Cc. Fraysse et al. [25] in a recent neutron powder diffraction study have also proposed the Cc space group for this composition range. However, recently several workers [26] have questioned the Cc space group below $T_{AFD}$ in this composition range. According to these workers, two ferroelectric phases in the space groups R3m & Cm and R3c & Cm coexist above and below $T_{AFD}$, respectively. This proposal implies that the AFD transition involves a change of space group from R3m to R3c. While the R3c to R3m AFD phase transition has been known and accepted for several decades [11, 27, 28], the origin and the role of the coexisting monoclinic phase in the Cm space group below and above $T_{AFD}$ remains unexplained. Why only the R3m phase undergoes an AFD transition and not the Cm phase, even though both are ferroelectric with untilted oxygen octahedra, is also not clear?

The present work was undertaken to resolve the existing controversies about the structure of the ground state phase of PZT on the $Zr^{4+}$-rich side in the composition range $0.620\lesssim x\lesssim 0.940$ using synchrotron x-ray powder diffraction, neutron powder diffraction and dielectric studies. We have analyzed $Zr^{4+}$-rich compositions of 6% $Sr^{2+}$-substituted PZT, i.e. $(Pb_{0.94}Sr_{0.06})(Zr_xTi_{1-x})O_3$ (PSZT), with x=0.65, 0.70, 0.80, 0.85 and 0.90 since $Sr^{2+}$ substitution decreases the average cationic radius of the 'A' site in the $ABO_3$ perovskite structure and therefore expected to increase the tilt angle [29, 30], and concomitantly the intensity of the superlattice peaks also [31]. Further, to establish the structure of high temperature phase of these compositions, we have chosen a representative composition of PSZT with x=0.65 and analyzed SXRPD patterns as a function of temperature. Our results unambiguously reject the possibility of R3m to R3c AFD transition with or without the coexisting monoclinic Cm phase. It is shown that the AFD transition occurs between a monoclinic phase in the Cm space group stable for $T_{AFD}<T<T_c$ and another monoclinic phase in the Cc space group below $T<T_{AFD}$, over the entire composition range $0.620\lesssim x\lesssim 0.940$. We also use the present results in conjunction with our previous results [32] to extend the phase diagram proposed earlier by us in the MPB region [32] to $Zr^{4+}$-richer side.

## II. Experimental and Analysis

Samples of PSZT for $0.65\leq x\leq 0.90$ were prepared by solid state reaction route. Stoichiometric mixtures of analytical reagent grade $PbCO_3$ (99%), $SrCO_3$ (99%), $ZrO_2$ (99%), and $TiO_2$ (99%) were calcined at 1073 K for 6 h in an open alumina crucible and sintered at 1323 K for 6 h in closed alumina crucibles using $PbZrO_3$ as spacer powder. The weight loss in the samples after sintering was $\lesssim$0.1%.

Temperature dependent synchrotron x-ray powder diffraction (SXRPD) measurements in the 300 to 620K range and also at room temperature were carried out at P02.1 hard x-ray diffraction beam line of PETRA III, DESY, Germany, at a wavelength of 0.20712Å (~60keV). Data were collected at a step of $0.007^0$ in the 2θ range from 1.5 to $11^0$. Neutron diffraction data at room temperature on the same samples were obtained from FRMII, Garching, Germany, using the high-resolution SPODI powder diffractometer [33] at a wavelength of 2.536 Å.

Rietveld analysis of the neutron and synchrotron x-ray powder diffraction patterns was carried out using FULLPROF [34] software package.

For dielectric measurements, fired-on silver electrodes were applied after gently cleaning the surfaces of the pellets using 0.25μm diamond paste. The temperature dependent dielectric measurements were carried out at a heating rate of 1K/min using a Novocontrol Alpha-A high performance frequency analyzer and a Eurotherm 2404 programmable temperature controller. The sample temperature during dielectric measurements was varied using a home made high temperature furnace and was controlled within ±1K using a Eurotherm-2404 programmable temperature controller.

## III. Results and discussion

In our previous work [22], samples were prepared by semi-wet route and it has been shown that the MPB in PSZT lies in the composition range $0.520\lesssim x\lesssim 0.535$ [Fig. 1 of ref. 22] whereas it lies in the range $0.515<x<0.530$ for pure PZT prepared by the same method [9, 17]. The physical properties show a peak at x=0.530 for PSZT with 6% Sr substitution [35, 36] whereas they peak at x=0.520 for pure PZT [11]. Thus, the MPB shifts by $\Delta x\simeq 0.01$ on 6% $Sr^{2+}$ substitution at $Pb^{2+}$-site. However, the symmetries of the phases across the MPB and in the MPB region remain exactly the same. This was shown in our earlier papers [22, 32, 37] for 6% Sr substitution. No new crystallographic phase emerges as a result of 6% Sr substitution. It is therefore not expected that the same materials prepared by solid state route will introduce new phases in the phase diagram of PSZT or PZT. Solid solution route can increase the width of the MPB region due to compositional homogeneities in the samples. This has been discussed in our early papers related to the location of MPB width for the semi-wet route [38]. Fig. 1 of supplementary file gives high resolution SXRPD collected at room temperature to confirm that our samples are single phase and they show the features identical to those for pure PZT, other than a shift of the MPB by $\Delta x\simeq 0.01$. With these remarks about our sample vis-à-vis PZT, we now proceed to describe phase transitions in PSZT.

**(a) Antiferrodistortive phase transition in PSZT compositions with $0.65 \leq x \leq 0.90$: Dielectric studies**

As mentioned earlier, the phase transitions on the $Zr^{4+}$-rich side of PZT are driven by both FD polar and AFD anti-phase rotational instabilities occurring at $T_c$ and $T_{AFD}$ with $T_{AFD}<T_c$. The ferroelectric or polar transition at $T_c$ gives rise to a huge peak in the real part of the dielectric permittivity. Fig. 1 depicts the temperature dependence of the real ($\varepsilon'$) and imaginary parts ($\varepsilon''$) of the dielectric permittivity for PSZT compositions with x=0.65, 0.70, 0.80, 0.85 and 0.90. For comparison, we have also included $\varepsilon'(T)$ plot for a typical tetragonal composition with x=0.40. It is evident from the figure that with the increase of $Zr^{4+}$-content, the Curie temperature ($T_c$) decreases. The sudden upward trend in the dielectric constant ($\varepsilon'$) above $T_c$ for some compositions is due to the increase in the conductivity of the samples at lower frequencies.

The $\varepsilon'(T)$ variation for $x \geq 0.65$ is not sharp but rather diffuse as compared to the relatively sharp $\varepsilon'(T)$ for the tetragonal composition with x=0.40 (see Fig. 1). It is interesting to note that the diffuseness of the $\varepsilon'(T)$ initially increases with increasing x but starts decreasing for $x \geq 0.850$. The ferroelectrics showing such diffuse phase transitions exhibit significant deviation from Curie-Weiss law and often exhibit relaxor ferroelectric behavior [39, 40]. For relaxor ferroelectrics the temperatures $T'_m$ and $T''_m$ corresponding to the peaks in $\varepsilon'(T)$ and $\varepsilon''(T)$ plots shift to higher temperature on increasing the frequencies of measurement. In addition, $T''_m$ is less than $T'_m$. However, the $\varepsilon'(T)$ data shown in Fig. 1 reveals that $T'_m=T''_m=T_c$ and it does not shift with increasing frequency of measurement. This suggests that the diffuse phase transition in PSZT on the $Zr^{4+}$-rich side of the MPB is of non-relaxor type, as has been reported in other disordered ferroelectrics like $Pb(Fe_{0.50}Nb_{0.50})O_3$ [41].

Such diffuse transitions have generally been explained in terms of local transition temperatures in different regions of the sample with respect to the average transition temperature $T_c$ due to statistical fluctuations in the composition [39, 40, 42, 43]. The dielectric measurements correspond to the superimposition of contributions from individual regions with their characteristic transition temperatures giving rise to an average Curie transition temperature $T_c$ with a Curie range of temperatures ($\Delta T_c$) over which the transition is spread out. Assuming a Gaussian type fluctuation in local $T_c$, it has been shown that the critical exponent for $\varepsilon'(T)$ would be 2 for such diffuse phase transitions [42]:

$$\frac{1}{\varepsilon'} - \frac{1}{\varepsilon'_m} = \frac{[T - T'_m]^2}{2\delta^2} \qquad (1)$$

where $2\delta$ is the width of the Gaussian and determines the extent of the diffuseness of the phase transition. However, in several systems, the critical exponent has been reported to lie in the range $1<\gamma<2$, where $\gamma=1$ corresponds to the ideal mean-field Curie-Weiss behavior. We have determined $\gamma$ as a function of composition using the following empirical relationship, which is a generalization of Eqn. 1 [43]:

$$\frac{1}{\varepsilon'} - \frac{1}{\varepsilon'_m} = \frac{[T - T'_m]^\gamma}{C} \qquad (2)$$

Fig. 2 shows the fits for the above relationship for PSZT with x= x=0.40, 0.65, 0.70, 0.80, 0.85 and 0.90. The value of $\gamma$ is found to be 1.47, 1.57, 1.91, 1.79, 1.69 and 1.58, respectively.

The physics of such diffuse transitions due to quenched (frozen-in) disorder has been discussed by Imry and Wortice [44] for first order phase transitions. This theory takes into account the concept of fluctuations in local transition temperatures due to frozen-in disorder, as was considered empirically in early literature on diffuse ferroelectric transitions leading to Eqns. 1 and 2 [42, 43]. The essential argument behind the Imry and Wortice [44] theory is that the correlation length and the transition depends on the local composition (i.e. the distribution of the frozen-in disorder) and because of that the transition can get smeared provided the interfacial energy associated with finite size transformed clusters with their own characteristics transition temperatures can be recovered from the chemical free energy gain due to the local paraelectric to ferroelectric phase transition.

In PSZT, we have dilute disorder at $Pb^{2+}$ site due to 6% $Sr^{2+}$ substitution and significant frozen-in disorder at the $Zr^{4+}$ site due to the substitution by $Ti^{4+}$. This is frozen-in disorder or quenched disorder as it cannot be annealed out. If quenched disorder can be annealed out, the $Ti^{4+}$ and $Zr^{4+}$ should be ordered giving rise to super structures. The PSZT samples used by us are the annealed powders obtained from crushing of sintered pellets. No evidence for $Zr^{4+}$ and $Ti^{4+}$ ordering is observed after annealing and therefore the disorder introduced by $Ti^{4+}$ substitution is random and quenched-in or frozen-in. In such disordered systems, statistical fluctuation in local composition is the main cause for the diffuse nature of ferroelectric phase transition. Since such a smearing of $\varepsilon'(T)$ has been reported in $Zr^{4+}$-rich compositions of PZT also [45], we believe that the smearing of $\varepsilon'(T)$ in PSZT shown in Fig. 1 is essentially due to the substitution of $Ti^{4+}$ at the $Zr^{4+}$ site of $PbZrO_3$. This is why the transition becomes more diffuse on moving away from the $PbZrO_3$ end with gradually increasing value of $\gamma$ from 1.58 for x=0.90 to 1.91 for x=0.70. The value of $\gamma$ for x=0.70 is close to 2 expected for Gaussian distribution of $T_c$. For further increase in $Zr^{4+}$ content, $\gamma$ again starts decreasing as for the MPB compositions near x=0.50, the concept of $Ti^{4+}$ as a substitutional disorder in the $Zr^{4+}$ lattice loses meaning. Microscopically, the difference in the ionic radii of $Zr^{4+}$ (r~0.72Å) and $Ti^{4+}$( r~0.605Å) leads to larger Zr-O bond lengths (~2.1Å) as compared to Ti-O (~1.9Å) bond lengths in PZT as confirmed by pair distribution function (PDF) analysis [46], density functional theory (DFT) calculations [47] and x-ray absorption fine structure (XAFS) experiments [48]. It has been found that the off-centre displacement of $Zr^{4+}$ in oxygen octahedral cage is less than that of the $Ti^{4+}$ in the ferroelectric phase. This difference may be responsible for the strong disorder effect due to substitution of $Ti^{4+}$ at $Zr^{4+}$ site.

The $\varepsilon'(T)$ plots do not reveal any signature of the AFD transition below $T_c$, except for one composition i.e., x=0.90, for which a step like change is observed at the AFD transition. A step like anomaly at the $T_{AFD}$ appears because of the coupling between the polar and AFD modes but is generally overwhelmed by the strong overlapping contribution from the polar mode [30]. Since the relative strength of the AFD mode increases while that of the polar mode decreases on approaching the $PbZrO_3$ end, the signature of AFD transition in $\varepsilon'(T)$ is not observed for compositions farther away from the $PbZrO_3$ end. Interestingly, the dielectric loss (tan$\delta$ or $\varepsilon''(T)$) gives clearer signature of the non-polar AFD transitions. This has been discussed in detail in the context of $SrTiO_3$ where a peak in, $\varepsilon''(T)$ at cubic to tetragonal AFD phase transition is observed around $T_{AFD}$~105K, without any corresponding anomaly in $\varepsilon'(T)$ (see e.g. ref. 49). The peak in $\varepsilon''(T)$ at $T_{AFD}$ has been attributed to the dynamics of domain walls which are affected by the AFD transition involving tilting of oxygen octahedra [30, 50]. The peak in ε″(T) shown in Fig. 1 is also due to the AFD transition, as has been reported by previous workers also in the context of PZT [11, 27]. However, the peak in ε″(T) at $T_{AFD}$ peak is rather broad and such broad peaks in dielectric loss may also occur due to extrinsic reasons like defects/impurities dynamics. The defect peaks disappear at very high frequencies. In our case, as evident from insets of Fig. 1, the peak in ε″(T) is observable even at 1MHz frequency. Further, the temperature corresponding to the anomaly in ε″(T) is independent of the measuring frequency, temperature rate and thermal history, and can therefore be regarded to be of intrinsic origin and linked with the AFD transition involving tilting of oxygen octahedra. The reason why we did not observe peak in ε″(T) x=0.80 was due to overwhelming contribution of the $\sigma_{AC}$ of the samples. It is evident from Fig. 1 that $T_{AFD}$ is above room temperature for all the compositions and is found to initially increase with increasing $Zr^{4+}$-content, reach a maximum for x≈0.85 and then start decreasing in agreement with similar observations for undoped PZT. It is worth mentioning here that recently Cordero et al. [30] have observed such diffuse anomalies in undoped PZT in the imaginary part of dielectric susceptibility ($\chi''$) and elastic compliance (s″) around the $T_{AFD}$ and they have established the temperature-composition (T-x) phase diagram of PZT. The temperature corresponding to the peak in ε″(T) for our samples of PSZT is in agreement with the disappearance of the superlattice peaks in SXRPD patterns as discussed below.

**(b) Structural evidence for antiferrodistortive phase transition: SXRPD studies**

In this section, we consider the evolution of SXRPD profiles of PSZT with x=0.65 as a function of temperature for a representative composition. Fig. 3 depicts the synchrotron x-ray powder diffraction profiles of $(3/2\ 1/2\ 1/2)_{pc}$ superlattice reflection and $(200)_{pc}$, $(220)_{pc}$, and $(222)_{pc}$ pseudocubic (pc) perovskite reflections as a function of temperature for $(Pb_{0.94}Sr_{0.06})(Zr_{0.65}Ti_{0.35})O_3$ (PSZT65). On increasing the temperature, the intensity of the $(3/2\ 1/2\ 1/2)_{pc}$ superlattice peak decreases and vanishes above ≃315K, in agreement with the $T_{AFD}$ obtained by dielectric studies discussed in the previous section. Although the superlattice peak vanishes above $T_{AFD}$≃315K, the asymmetric nature of $(222)_{pc}$ reflection still persists above $T_{AFD}$. On increasing the temperature further, all the profiles eventually become singlets and symmetric, as expected for the cubic paraelectric phase.

For the AFD phase transitions in PZT, the tilt angle corresponds to the primary order parameter. The intensity of superlattice peaks is proportional to the square of tilt angle [31]. We depict in Fig. 4(a) the temperature dependence of the integrated intensity of the $(3/2\ 1/2\ 1/2)_{pc}$ superlattice peak for PSZT65 to determine the order of the AFD phase transition. The solid line in Fig. 4(a) represents a least-squares fit for the power law $I_{(3/2\ 1/2\ 1/2)pc} \sim (1-T/T_c)^{2\beta}$ which gives an exponent of 1/2. This suggests second order nature of the AFD phase transition in PSZT65. AFD phase transition in several well known perovskites like $SrTiO_3$ [51] and $CaTiO_3$ [52] is also known to be second order. Second order nature of the AFD phase transition in PSZT65 was further confirmed by the variation of unit cell volume with temperature depicted in Fig. 4(b). The continuous nature of volume with temperature with only a change of the slope around 315K is consistent with the second order nature of the AFD phase transition in PSZT65. We also observe a sharp change of slope in the temperature variation of the FWHM of $(200)_{pc}$ peak (Fig. 4(c)) around 315K corresponding to the AFD transition at $T_{AFD}$≃315K.

As a next step, we now consider the space group symmetry of the ferroelectric phase of PSZT65 in the temperature range $T_{AFD}<T<T_c$. In the early literature, this phase has been labeled as $F_R^{HT}$ phase with R3m space group, identical to the structure of PZT at room temperature for 0.530<x<0.620 [11]. It is now well established [9, 10] that the room temperature structure of PZT for 0.530<x<0.620 is not rhombohedral but monoclinic in the Cm space group. Accordingly, we considered both the R3m and Cm space group as plausible models for PSZT65 at $T_{AFD}<T<T_c$. Fig. 5 shows the Rietveld fits using these two space group models at T=360K. The Cm space group model gives significantly better fit (see Fig. 5(a) and (b)) and lower agreement factors compared to the R3m space group model. Following the results of ref. 26, we considered the coexistence of Cm and R3m phases but this led to zero phase fraction of the R3m phase. We are thus led to conclude that the structure of PSZT65 in the temperature range $T_{AFD}<T<T_c$ can be described using monoclinic Cm space group alone. We believe that the phase coexistence reported [26] for compositions corresponding to the Zr-rich side of the MPB with $T_{AFD}$ above room temperature is of extrinsic origin coming from chemical inhomogeneity in the samples as discussed in the literature for PZT samples prepared by different routes [53, 54].

**(c) Room temperature structures of PSZT for 0.65≤x≤0.90: SXRPD and neutron powder diffraction studies**

The intensity of superlattice peak increases as the $Zr^{4+}$-content increases which implies increase in the tilt angle [31]. This is illustrated in Fig. 1 of the supplementary file which depicts the SXRPD patterns for $(200)_{pc}$, $(220)_{pc}$ and $(222)_{pc}$ perovskite peaks along with the most intense superlattice peak with indices $(3/2\ 1/2\ 1/2)_{pc}$ at room temperature for various compositions of PSZT in the composition range of 0.65≤x≤0.90. *T*he ratio of FWHM
of the (200)pc peak to that of the nearby (111)pc peak for the tetragonal and rhombohedral structures should be ~1.0. However, we find that this ratio as can be seen from Fig. 2 of the supplementary file. This type of anisotropic broadening has been shown to be a signature of monoclinic distortion with short correlation length in the context of PZT as it cannot be modeled in terms of anisotropic strain using Stephen's model [54]. This implies that the room temperature structure of $Zr^{4+}$-rich compositions of PSZT for 0.650≤x≤0.90 may not be rhombohedral. Accordingly, we considered four different structural models for the refinement of the structure of PSZT65: (a) R3c space group without anisotropic strain broadening parameters, (b) R3c space group with anisotropic strain broadening parameters [55], (c) Cc space group [17, 25] and (d) R3c+Cm [26] space group models proposed in the literature.

Fig. 6 depicts the Rietveld fits for the four structural models. We first consider R3c space group model without anisotropic strain broadening parameters. The fits are shown in Fig. 6(a). This structural model is not able to account for the additional broadening of $(200)_{pc}$ peak. Also, the intensity of the $(200)_{pc}$, $(220)_{pc}$ and $(222)_{pc}$ peaks is not accountable with this model. Further, this space group model gives very high value of $\chi^2$ (17.00) and $R_{wp}$ (4.57). We then considered R3c space group model with anisotropic strain broadening parameters. It is evident from the difference profile (bottom line) of Fig. 6(b) that the R3c space group model with anisotropic strain broadening parameters improves the fits and reduces the $\chi^2$ (11.30) significantly but the mismatch between the observed and calculated peak profiles still persists. The Cc space group model, proposed by Pandey et al. [17] and Fraysse et al. [25], on the other hand, gives excellent fit between the observed and calculated profiles of the superlattice as well as the main perovskite reflections with drastically lower value of $\chi^2$ (3.70) in comparison to the R3c space group model (see Fig. 6(c)). Rietveld fits corresponding to the coexistence of R3c and Cm space group model proposed recently by some workers [26] for $Zr^{4+}$-rich compositions of PZT are shown in Fig. 6(d). Evidently, the R3c+Cm model also gives significantly higher value of $\chi^2$ (7.01) than the Cc space group model even though the latter has lesser number of refinable parameters (35) in comparison to the former (43). We can, therefore, rule out the R3c and R3c+Cm space group models proposed in ref.26 for the room temperature superlattice phase of PZT for $Zr^{4+}$-rich compositions. Combining the results of the previous and the present section, we thus conclude that the AFD transition occurs between two monoclinic phases in the Cm and Cc space groups stable above and below $T_{AFD}$, respectively.

On the same basis as discussed above for PSZT65, we carried out refinements of room temperature SXRPD patterns for other $Zr^{4+}$-rich compositions using the Cc space group model and they are shown in Fig. 7. The fits for all the compositions are quite good with nearly flat difference profiles. The intensity of the superlattice peaks resulting from the tilting of oxygen octahedra is rather small in Fig. 1 of supplementary file due to relatively weak scattering of x-rays from lighter atoms such as oxygen. Neutrons, on the other hand, are scattered by the nucleus of atoms and neutron scattering length does not vary with atomic number in a regular manner. Because of this, light elements such as oxygen etc. also produce relatively strong scattering. So, to complement our SXRPD studies, we refined the structure of PSZT using neutron powder diffraction patterns also for two compositions with x=0.65 (PSZT65) and x= 0.70 (PSZT70). We considered the same structural models as in the refinements using SXRPD data. The Rietveld fits corresponding to R3c and R3c+Cm space group models reveal significant mismatch between the observed and calculated profiles of $(3/2\ 1/2\ 1/2)_{pc}$ superlattice peak. The calculated profile for this peak is not able to capture the width of the observed profile (Fig. 8 and 9 (a) and (b)). This mismatch is highlighted with arrows in the figure. The Cc space group model gives much better fit in comparison to R3c and R3c+Cm models to the superlattice $(3/2\ 1/2\ 1/2)_{pc}$ peak with lowest $\chi^2$ values for both the compositions (see Fig. 8(c)). We can therefore rule out the R3c and R3c+Cm structural models for PSZT65 and 70 using Rietveld analysis of the neutron powder diffraction data also. The first row of Fig. 8 and 9 depict the Rietveld fits around the $(1/2\ 1/2\ 1/2)_{pc}$ superlattice position which is expected for the Cc space group model but extinguished for the R3c and R3c+Cm models [26]. The intensity of this peak is about ~0.02% of the highest intensity $(111)_{pc}$ perovskite peak of the neutron powder diffraction pattern whereas the background noise level is ~0.03%. We therefore do not expect to resolve the $(1/2\ 1/2\ 1/2)_{pc}$ peak in our neutron diffraction pattern. Based on our Rietveld refinements using SXRPD and neutron powder diffraction patterns, we thus conclude that the ground state of PSZT in the composition range $0.620 \lesssim x \lesssim .940$ and below $T_{AFD}$ corresponds to a monoclinic phase in the Cc space group confirming the proposal of Pandey et al. [17].

**(d) Phase diagram of PSZT for $0.40 \leq x \leq 0.90$**

On the basis of the high resolution SXRPD, neutron powder diffraction, dielectric and sound velocity studies carried out in our previous and in the present work, we have constructed a temperature (T)-composition (x) phase diagram of PSZT for $0.40 \leq x \leq 0.90$ shown in Fig. 10. Compositions in the range $0.51 \leq x \leq 0.550$ were prepared by semi-wet route. Full description of sample preparation has been given in our previous work [22]. The paraelectric to ferroelectric phase boundary (red dots) for compositions with $0.51 \leq x \leq 0.550$ were determined from the temperature dependent dielectric studies during heating cycles [32]. Dielectric studies for these compositions show vanishing of thermal hysteresis for $x \simeq 0.550$ [32] due to a crossover from first order to second order phase transition at a tricritical point at $x \simeq 0.550$. Using Rietveld analysis of temperature dependent SXRPD studies, it was shown that compositions with $0.515 \leq x \leq 0.545$ undergo cubic to tetragonal phase transition and then transform to pseudotetragonal monoclinic phase while the compositions with $0.550 \lesssim x \lesssim 0.940$ directly transform to the pseudorhombohedral monoclinic phase. This leads to a triple point in the phase diagram at $x \simeq 0.550$ coinciding with tricritical point [32]. We show in Fig. 10 the magnified view of this region to highlight the various phases in and around the tricritical point. The low temperature ferroelectric to antiferrodistortive phase boundary (black dots) for the compositions with $0.515 \leq x \leq 0.550$ was established using temperature dependent sound velocity measurements. The symmetries below the $T_{AFD}$ were confirmed to be monoclinic in Cc space group [22, 37] except for x=0.515. The space group of the low-temperature phase of PSZT515 may be I4cm or Cc [17, 23] but more work is needed to settle this issue. The vertical dotted lines in Fig. 10 are the first order lines which separate the tetragonal and pseudorhombohedral monoclinic phases. MPB in PSZT lies for $0.520 \lesssim x \lesssim 0.535$ [22]. For $0.520 \lesssim x \lesssim 0.525$, pseudotetragonal monoclinic phase in Cm space group and tetragonal phase in P4mm space group are the stable phases while for $0.525 \leq x \lesssim 0.535$ pseudotetragonal and pseudorhombohedral monoclinic phases both in Cm space group coexist. According to Mishra [35] and Ramjilal et al. [36], the maximum in physical properties such as electromechanical coupling coefficient, piezoelectric strain coefficient and dielectric constant occurs for x=0.530. Thus MPB in PSZT is shifted by ~1% to Zr-rich side in comparison to undoped PZT where the maximum in physical properties occurs at $x \simeq 0.520$. Ti-rich compositions of PZT undergo only a high temperature paraelectric cubic to ferroelectric tetragonal phase transition. We have determined the Curie temperature ($T_c$) in this composition range using dielectric measurements.

Phase boundary for the composition range $0.65 \leq x \leq 0.90$ with $T_{AFD}$ lying above room temperature (300K) has been determined using anomalies in dielectric loss shown in Fig. 1. It is evident that $T_{AFD}$ determined using sound velocity measurements below room temperature for $x<0.620$ and those determined using dielectric measurements above room temperature for $0.620<x<0.940$ follow same trend as in case of pure PZT [11, 17]. It increases on increasing the Zr-content up to $x \gtrsim 0.850$ nearly linearly and then decreases. Ferroelectric $T_c$ for Zr-rich compositions with $0.650 \leq x \leq 0.90$ have been determined from the real ($\varepsilon'$) part of dielectric constant discussed in section III (a) and shown in Fig. 1. The symmetries of the various phases above and below $T_{AFD}$ as discussed in section III (b and c) have been determined using SXRPD and neutron powder diffraction studies. For $T<T_{AFD}$ symmetry of Zr-rich compositions is monoclinic in Cc space group while for $T_{AFD}<T<T_c$, the symmetry is also monoclilinic but in Cm space group. All the compositions transform to the cubic paraelectric phase in $Pm\bar{3}m$ space group above their corresponding Curie temperature ($T_c$) and $T_c$ decreases with increasing $PbZrO_3$ content, as expected from the lower $T_c$ of $PbZrO_3$. Here, it is important to emphasize that 6% $Sr^{2+}$ substitution does not alter the phase stabilities of PZT significantly and the PSZT phase diagram is topologically similar to that of pure PZT.

## IV. Conclusions

This study has led to the several important findings regarding the nature of AFD phase transition on the $Zr^{4+}$-rich side of PZT and its phase diagram. Firstly, we have resolved the existing controversies about the structures of PZT above and below the $T_{AFD}$ for the composition range $0.620 \lesssim x \lesssim 0.940$ using 6% $Sr^{2+}$-substituted samples. Only the Cc space group model with lesser number of refinable parameters in comparison to R3c [11] or R3c+Cm [26] structural models gives lower values of $\chi^2$ and $R_{wp}$ and can also account for the width of the $(3/2\ 1/2\ 1/2)_{pc}$ superlattice peak in the neutron powder diffraction patterns. Our results thus rule out the R3c and R3c+Cm structural models proposed in recent years [26] and confirm that Cc space group model represents the true structure of the ground state phase of PZT at $T<T_{AFD}$ for the composition range $0.620 \lesssim x \lesssim 0.940$. Also, the temperature dependence of volume and integrated intensity for PSZT65 shows that AFD phase transition is of second order type. Secondly, we have also shown that the high temperature ferroelectric phase at $T_{AFD}<T<T_c$ has got Cm space group and not R3m or R3m+Cm as proposed in several recent studies [26]. Thirdly, the phase diagram of PSZT for the composition range $0.40 \leq x \leq 0.90$ has been established using neutron powder diffraction, synchrotron x-ray powder diffraction, dielectric and sound velocity studies of the present and previous works [22, 32, 37]. In contrast to the old phase diagram of PZT given in the standard texts [11], it is shown that all the compositions of PSZT with $0.515 \lesssim x \lesssim 0.90$ exhibit the AFD phase transition involving two monoclinic phases in the Cm and Cc space groups stable above and below $T_{AFD}$, respectively. The $T_{AFD}$ of the compositions in the range $0.515 \lesssim x \lesssim 0.62$ lies below room temperature while for $0.62<x \lesssim 0.90$, it lies above the room temperature. Because of the tilted nature of MPB, a P4mm to Cm polymorphic phase transition precedes the Cm to Cc AFD phase transition for the composition range $0.515<x \lesssim 0.550$. The present work along with our previous findings [9, 10, 17] show that the monoclinic Cm phase and not R3m phase is stable above the $T_{AFD}$ line in the entire T-x plane on the $Zr^{4+}$-rich side of the MPB. Our results thus confirm the phase diagram proposed by Pandey et al. [17] for undoped PZT and reject the alternative structural models below and above $T_{AFD}$ [26, 28, 55] and alternative phase diagrams proposed in some recent studies [11, 23, 26].

D. Pandey acknowledges financial support from Science and Engineering Research Board (SERB) of India through J. C. Bose National Fellowship grant. R. S. Solanki acknowledges financial support from SERB, India in the form of Research Assistantship under J. C. Bose Fellowship. We acknowledge the assistance of Dr. M. Hinterstein from PETRA III, DESY, Germany in collection of SXRPD Data. Authors RSS and DP acknowledge the financial support from Saha Institute of Nuclear Physics (SINP) under the DST-DESY project to carry out SXRPD experiments at DESY, Germany.

## References

[1] R. Blinc and B. Zeks, Soft Modes in Ferroelectrics and Antiferroelectrics, North Holland Publishing Co.-Amsterdam, Oxford, p. 1, 1974.

[2] A. D. Bruce and R. A. Cowley, Structural Phase Transitions III: Critical dyanamics and quasi elastic scattering, Adv. in Phys., **29,** 219 (1980).
[3] M. E. Lines and A. M. Glass, Principles and Applications of Ferroelectrics and Related Materials, Clarendon Press, Oxford, p. 241, 1977.
[4] H. Unoki and T. Sakudo, J. Phys. Soc. Japan., **23,** 546 (1967).
[5] P. A. Fleury, J. F. Scott, and J. M. Worlock, Phys. Rev. Lett. **21**, 16 (1968).
[6] R. A. Cowley, W. J. L. Buyers and G. Dolling, Solid State Comm., **7,** 181 (1969).
[7] Sanjay Kumar Mishra and Dhananjai Pandey, Appl. Phys. Lett. **95**, 232910 (2009).
[8] Jay Prakash Patel, Anar Singh and Dhananjai Pandey, J. Appl. Phys. **107**, 104115(2010); Anar Singh, Anatoliy Senyshyn, Hartmut Fuess and Dhananjai Pandey, J. Appl. Phys. **10**, 024111 (2011).
[9] Ragini, R. Ranjan, S. K. Mishra, and D. Pandey, J. Appl. Phys. **92**, 3266 (2002).
[10] A. K. Singh, D. Pandey, S. Yoon, S. Baik, and N. Shin, Appl. Phys. Lett. **91**, 192904 (2007).
[11] B. Jaffe, W. R. Cook and H. Jaffe, Piezoelectric Ceramics, Academic Press, London, p. 135, 1971.
[12] S. K. Mishra, D. Pandey, and A. P. Singh, Appl. Phys. Lett. **69**, 1707 (1996).
[13] S. K. Mishra, A. P. Singh, and D. Pandey, Philosophical Magazine **76**, 213 (1997).
[14] B. Noheda, D. E. Cox, G. Shirane, J. A. Gonzalo, L. E. Cross and S. E. Park, Appl. Phys. Lett. **74**, 2059 (1999); B. Noheda, J. A. Gonzalo, L. E. Cross, R. Guo, S.-E. Park, D. E. Cox, and G. Shirane, Phys. Rev. **B 61,** 8687 (2000).
[15] Ragini, S. K. Mishra, D. Pandey, H. Lemmens and G. V. Tendeloo, Phys. Rev. **B 64**, 054101 (2001).
[16] R. Ranjan, Ragini, S. K. Mishra, D. Pandey, and B. J. Kennedy, Phys. Rev. **B 65**, 060102(R) (2002).
[17] D. Pandey, A. K. Singh, and S. Baik, Acta Crystallogr. Sect. **A 64**, 192 (2008).
[18] D. M. Hatch, H. T. Stokes, R. Ranjan, Ragini, S. K. Mishra, D. Pandey and B. J. Kennedy, Phys. Rev. **B 65**, 212101 (2002).
[19] D. E. Cox, B. Noheda, and G. Shirane, Phys. Rev. **B 71**, 134110 (2005).
[20] D. I. Woodward, J. Knudsen, and I. M. Reaney, Phys. Rev. **B 72**, 104110 (2005).
[21] J. Rouquette, J. Haines, V. Bornand, M. Pintard, Ph. Papet, W. G. Marshall, and S. Hull, Phys. Rev. **B 71**, 024112 (2005).
[22] R. S. Solanki, A. K. Singh, S. K. Mishra, S. J. Kennedy, T. Suzuki, Y. Kuroiwa, C. Moriyoshi, and D. Pandey, Phys. Rev. **B 84,** 144116 (2011).
[23] I. A. Kornev, L. Bellaiche, P. E. Janolin, B. Dkhil, and E. Suard, Phys. Rev. Lett. **97**, 157601 (2006).
[24] R. S. Solanki, S. K. Mishra, A. Senyshyn, S. Yoon, S. Baik, N. Shin, and D. Pandey, Appl. Phys. Lett. **102**, 052903 (2013).
[25] Guillaume Fraysse, Julien Haines, Veronique Bornand, Jerome Rouquette, Marie Pintard, Philippe Papet, and Steve Hull, Phys. Rev. **B 77,** 064109 (2008).
[26] H. Yokota, N. Zhang, A. E. Taylor, P. A. Thomas, and A.M. Glazer, Phys. Rev. **B 80**, 104109 (2009); D. Phelan, X. Long, Y. Xie, Z.-G. Ye, A. M. Glazer, H. Yokota, P. A. Thomas, and P. M. Gehring, Phys. Rev. Lett. **105**, 207601 (2010); N. Zhang, H. Yokota, A. M. Glazer, and P. A. Thomas, Acta Crystallogr. Sect. **B 67**, 386 (2011).
[27] H. M. Barnett, J. Appl. Phys., **33,** 1606 (1962); D. Berlincourt, H. H. A. Krueger and B. Jaffe, J. Phys. Chem. Solids **25,** 659-674 (1964).
[28] C. Michel, J. M. Moreau, G. D. Achenbach, R. Gerson and W. J. James, Solid State Commun., **7,** 865-868, (1969).
[29] J. B. Goodenough, Rep. Prog. Phys. **67,** 1915–93, (2004).
[30] F. Cordero, F. Craciun, and C. Galassi, Phys. Rev. Lett. **98,** 255701

(2007); F. Cordero, F. Trequattrini, F. Craciun, and C. Galassi, Phys. Rev. **B 87,** 094108, (2013).
[31] A. M. Glazer, Acta Crystallogr. Sect. B **28**, 3384 (1972); Acta Crystallogr. Sect. A **31**, 756 (1975).
[32] Ravindra Singh Solanki, S. K. Mishra, Chikako Moriyoshi, Yoshihiro Kuroiwa and Dhananjai Pandey, Phys. Rev. B, **88,** 184109 (2013).
[33] M. Hoelzel, A. Senyshyn, N. Juenke, H. Boysen, W. Schmahl, and H. Fuess, Nucl. Instrum. Methods Phys. Res. **A 667,** 32 (2012).
[34] J. Rodriguez-Carvajal Laboratory, FULLPROF, a Rietveld and pattern matching and analysis program, version July 2011, Laboratoire Leon Brillouin, CEA-CNRS, France; J. Rodriguez-Carvajal, Physica **B 192,** 55 (1993).
[35] S. K. Mishra, Studies on morphotropic phase boundary in PZT ceramics, Ph.D. thesis, Institute of Technology, Banaras Hindu University, 1998.
[36] R. Lal, R. Krishnan, and P. Ramakrishnan, Trans. Brit. Ceram. Soc. **87**, 99 (1988).
[37] R. S. Solanki, S. K. Mishra, A. Senyshyn, I. Ishii, C. Moriyoshi, T. Suzuki, Y. Kuroiwa, and D. Pandey, Phys. Rev. **B 86,** 174117 (2012).
[38] A. P. Singh, S. K. Mishra, D. Pandey, Ch. D. Prasad, and R. Lal, J. Mater. Sci. 28, 5050 (1993).
[39] G. A. Smolenskii, J. Phys. Soc. Japan, **28,** suppl., 26-37 (1970).
[40] L. E. Cross, Ferroelectrics **76**, 241 (1987); L. E. Cross, Ferroelectrics **151,** 305 (1994).
[41] S. P. Singh, A. K. Singh, D. Pandey, H. Sharma, O. Parkash, J. Mater. Res. **18,** 2677, (2003).
[42] V.V. Kirillov, Ferroelectrics **5,** 3 (1973).
[43] K. Uchino and S. Namura, Ferroelectrics Lett. **44,** 56 (1982).
[44] Y. Imry and M. Wortis, Phys. Rev. B **19,** 3580 (1979).
[45] I. Franke et al., J. Phys. D: Appl. Phys. **38,** 749-753 (2005).
[46] W. Dmowski, T. Egami, L. Farber, and P. K. Davies, in Fundamental Physics of Ferroelectrics 2001, edited by Henry Krakauer, AIP Conf. Proc.No. 582 (AIP, Melville, NY, 2001), p. 33.
[47] I. Grinberg, V. R. Cooper, and A. M. Rappe, Nature (London) **419**, 909 (2002).
[48] D. Cao, I.-K. Jeong, R. H. Heffner, T. Darling, J.-K. Lee, F. Bridges, J.-S. Park and K.-S. Hong, Phys. Rev. B **70**, 224102 (2004).
[49] R. Viana, P. Lunkenheimer, J. Hemberger, R. Böhmer, and A. Loidl, Phys. Rev. B **50,** 601(R) (1994).
[50] K. Fossheim and B. Berre, Phys. Rev. **B 5,** 3292 (1972).
[51] M. A. Geday and A. M. Glazer, J. Phys.: Condens. Matter **16,** 3303 (2004).
[52] Simon A. T. Redfern, J. Phys.: Condens. Matter **8,** 8267 (1996).
[53] P. Arigur and L. Benguigui, J. Phys., D., **8,** 1856 (1975); K. Kakegawa, J. Mohri, K. Takahashi, H. Yamamua and S. Shirashki, Solid St. Commun., **24,** 769 (1977); A. P. Wilkinson, J. Xu, S. Pattanaik, and S. J. L. Billinge, Chem. Mater. **10**, 3611 (1998).
[54] P. W. Stephens, J. Appl. Crystallogr. **32**, 281 (1999).
[55] A. M. Glazer and S. A. Mabud, Acta. Cryst. **B 34,** 1060 (1978).

## Figure Captions

**Fig. 1.** (Color online) Temperature dependence of real (ε′) and imaginary (ε″) part of the dielectric permittivity of PSZT for x= (a) 0.40, (b) 0.65, (c) 0.70, (d) 0.80, (e) 0.85 and (f) 0.90 at various frequencies (10 kHz, 50 kHz, 100 kHz, 500 kHz and 1 MHz). Insets show the variation of ε″ with temperature at 1 MHz frequency in the selected temperature range.

**Fig. 2.** (Color online) Temperature dependence of $[1/\varepsilon'-1/\varepsilon'_m]$ for PSZT for x=(a) 0.40, (b) 0.65, (c) 0.70, (d) 0.80, (e) 0.85 and (f) 0.90 using dielectric data at 1 MHz. The continuous dark line is the fit to the modified Curie–Weiss law [Eq. (2) in the text] with an exponent $1<\gamma<2$.

**Fig. 3.** (Color online) Temperature evolution of synchrotron XRPD profiles of the $(3/2\ 1/2\ 1/2)_{pc}$ superlattice reflection and $(200)_{pc}$, $(220)_{pc}$, and $(222)_{pc}$ perovskite reflections of PSZT65.

**Fig. 4.** (Color online) Temperature dependence of (a) integrated intensity, (b) unit cell volume and (c) FWHM of $(200)_{pc}$ peak in the selected temperature region. Continuous line in (a) represent the least square fit using power law $I_{(3/21/2/1/2)pc}$~$(1-T/T_c)^{2\beta}$.

**Fig. 5.** (Color online) Observed (dots), calculated (continuous line), and difference (bottom line) profiles of PSZT65 at 360K obtained after Rietveld refinement using (a) R3m and (b) Cm structural models. Vertical ticks below the peaks mark position of the Bragg reflections.

**Fig. 6**. (Color online) The observed (dots), calculated (continuous line) and difference (bottom) SXRPD profiles for the $(3/2\ 1/2\ 1/2)_{pc}$ superlattice and $(200)_{pc}$, $(220)_{pc}$, $(222)_{pc}$ perovskite peaks obtained after Rietveld refinement of the structure of PSZT65 at room temperature using (a) R3c space group without anisotropic strain broadening parameters, (b) R3c space group with anisotropic strain broadening parameters, (c) Cc space group and (d) R3c+Cm space group models.

**Fig. 7.** (Color online) Observed (open circles), calculated (solid line), and difference (bottom line) patterns obtained from the Rietveld analysis of the room temperature synchrotron XRPD data of PSZT70, 80, 85 and 90 using Cc space group model. The vertical tick marks above the difference line stand for the Bragg peak positions.

**Fig. 8**. (Color online) The observed (dots), calculated (continuous line) and difference (bottom) neutron diffraction profiles for PSZT65 for the $(3/2\ 1/2\ 1/2)_{pc}$ , $(3/2\ 3/2\ 1/2)_{pc}$ and $(5/2\ 1/2\ 1/2)_{pc}$ superlattice peaks obtained after Rietveld refinement of the structure at room temperature using (a) R3c, (b) Cc and (c) R3c+Cm space group models. First row of the figure shows the zoomed view around $(1/2\ 1/2\ 1/2)_{pc}$ superlattice reflection.

**Fig. 9**. (Color online) The observed (dots), calculated (continuous line) and difference (bottom) neutron diffraction profiles for PSZT70 for the $(3/2\ 1/2\ 1/2)_{pc}$ , $(3/2\ 3/2\ 1/2)_{pc}$ and $(5/2\ 1/2\ 1/2)_{pc}$ superlattice peaks obtained after Rietveld refinement of the structure at room temperature using (a) R3c, (b) Cc and (c) R3c+Cm space group models. First row of the figure shows the zoomed view around $(1/2\ 1/2\ 1/2)_{pc}$ superlattice reflection.

**Fig. 10.** (Color online) Schematic phase diagram of PSZT constructed on the basis of neutron, synchrotron, sound velocity and dielectric measurements. Red Dots in the phase diagram correspond to the high temperature cubic ($Pm\bar{3}m$) to tetragonal (P4mm) and cubic to pseudorhmbohedral monoclinic (Cm) phase transition temperatures during heating, obtained from dielectric measurements. The black dots correspond to low temperature antiferrodistortive ($T_{AFD}$) phase (space group Cc) transition temperatures. For $0.515 \leq x \leq 0.550$, $T_{AFD}$ was determined from the sound velocity measurements [32] while for $0.65 \leq x \leq 0.90$, it was determined from the temperature dependence of ε''. The symbols PT and PR represent pseudotetragonal and pseudorhombohedral monoclinic phases, respectively. The vertical lines are first order phase boundaries across which the two neighbouring phases

coexist. Also, the inclined P4mm-Cm (PT)/Cm (PR) phase boundary is a first order phase boundary. $T_c$ stands for Curie temperature and $T_{AFD}$ for antiferrodistortive phase transition temperature. Inset show the magnified view of figure around the tricritical point established in ref. 32.

# Figures

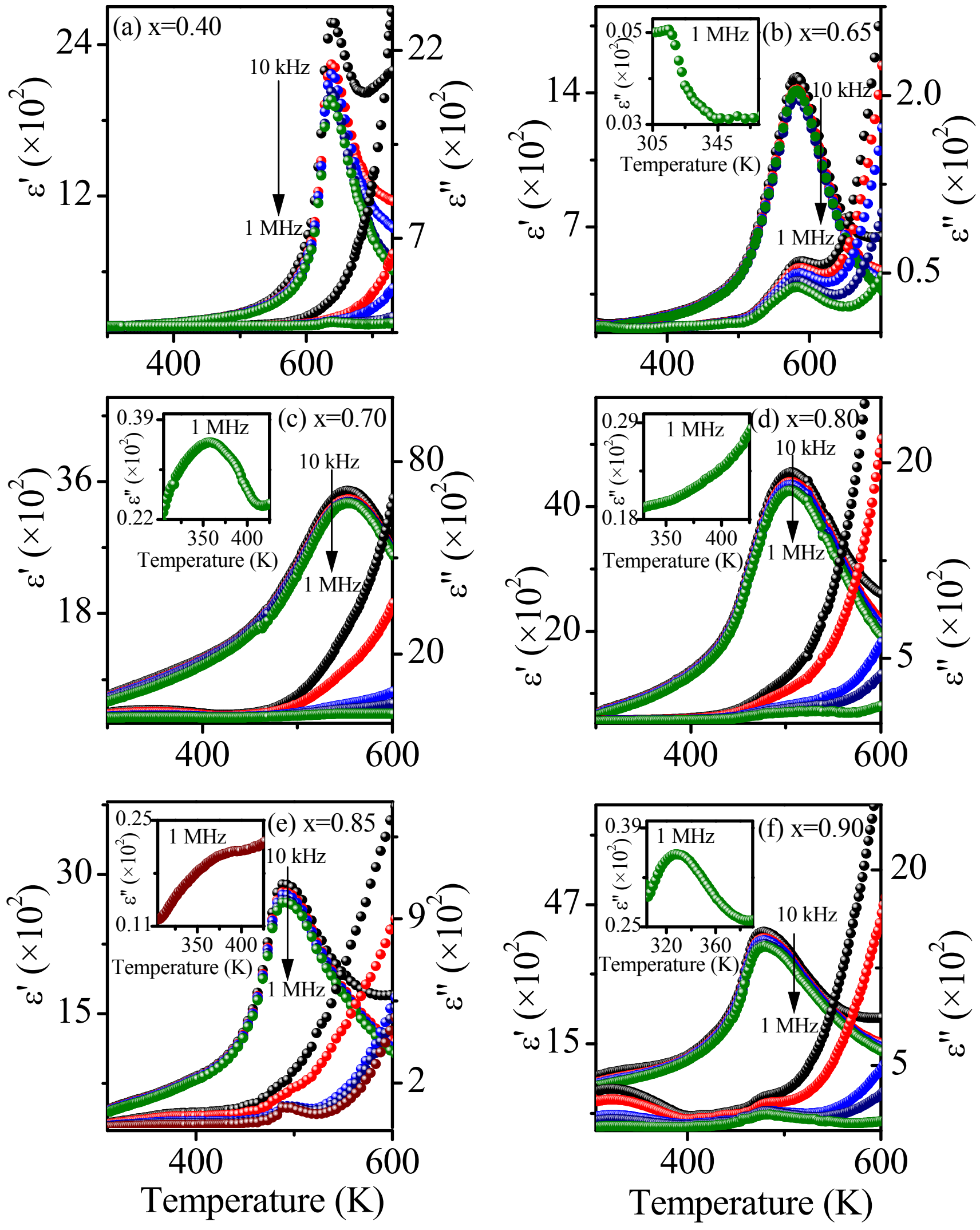

Fig.1

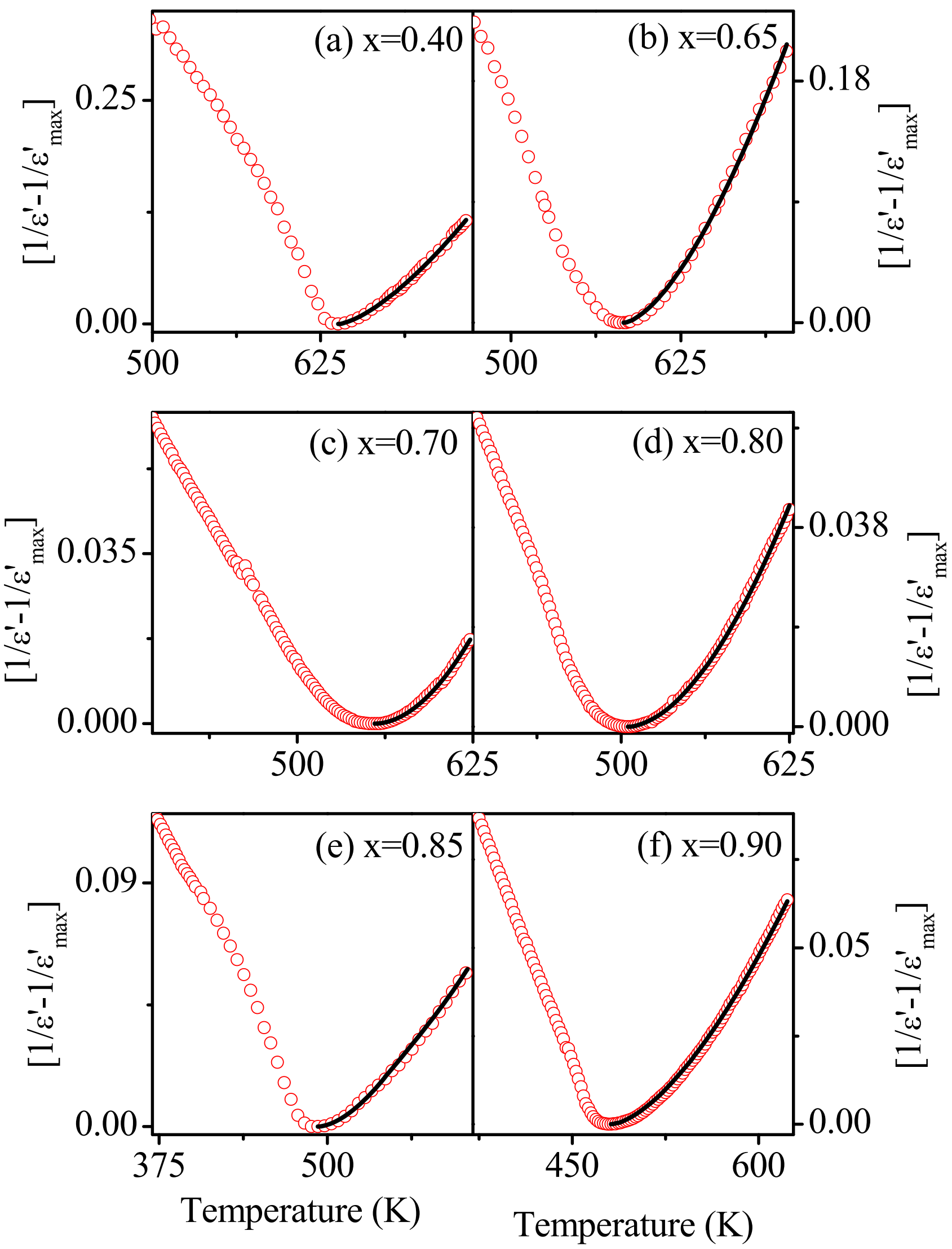

Fig. 2

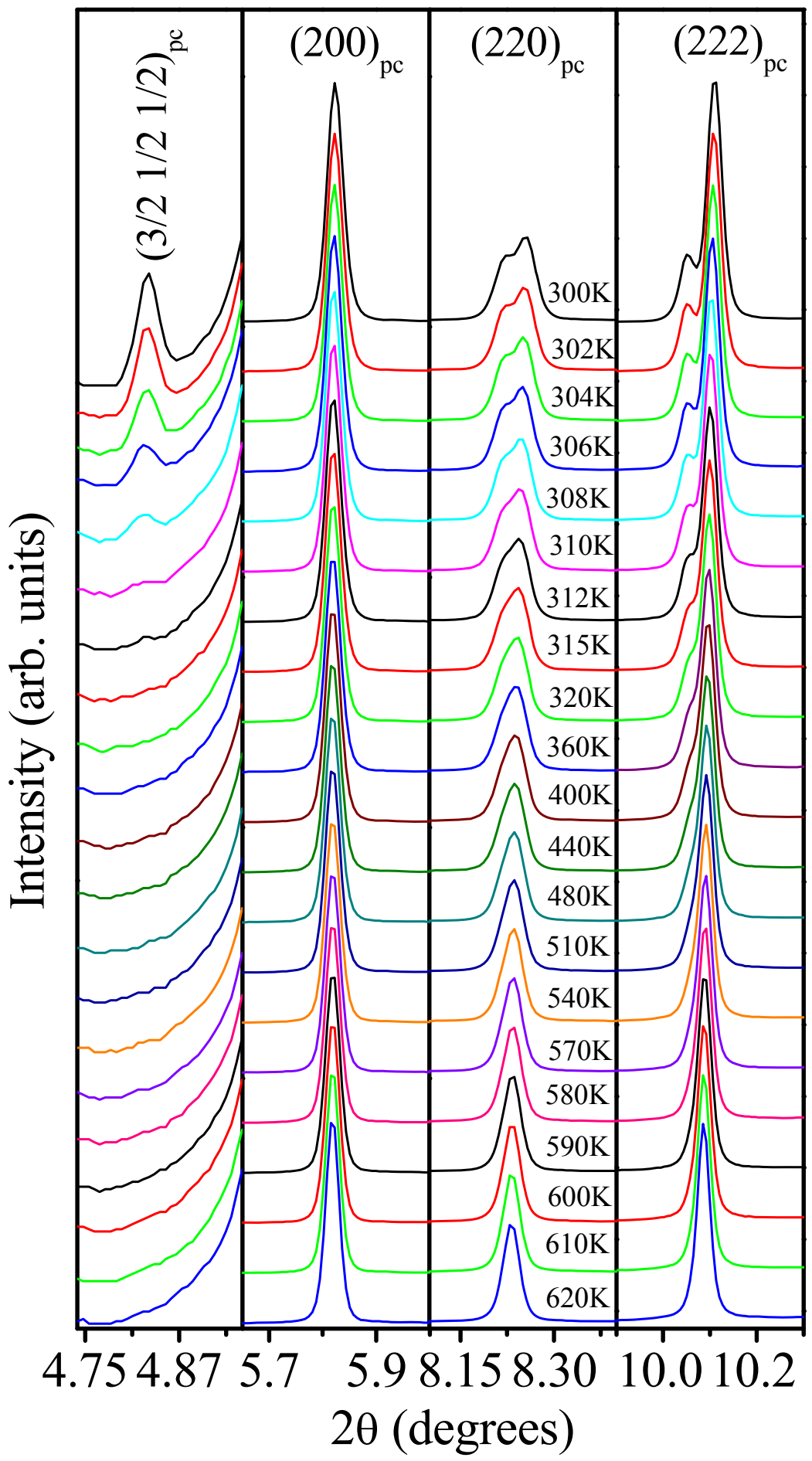

Fig. 3

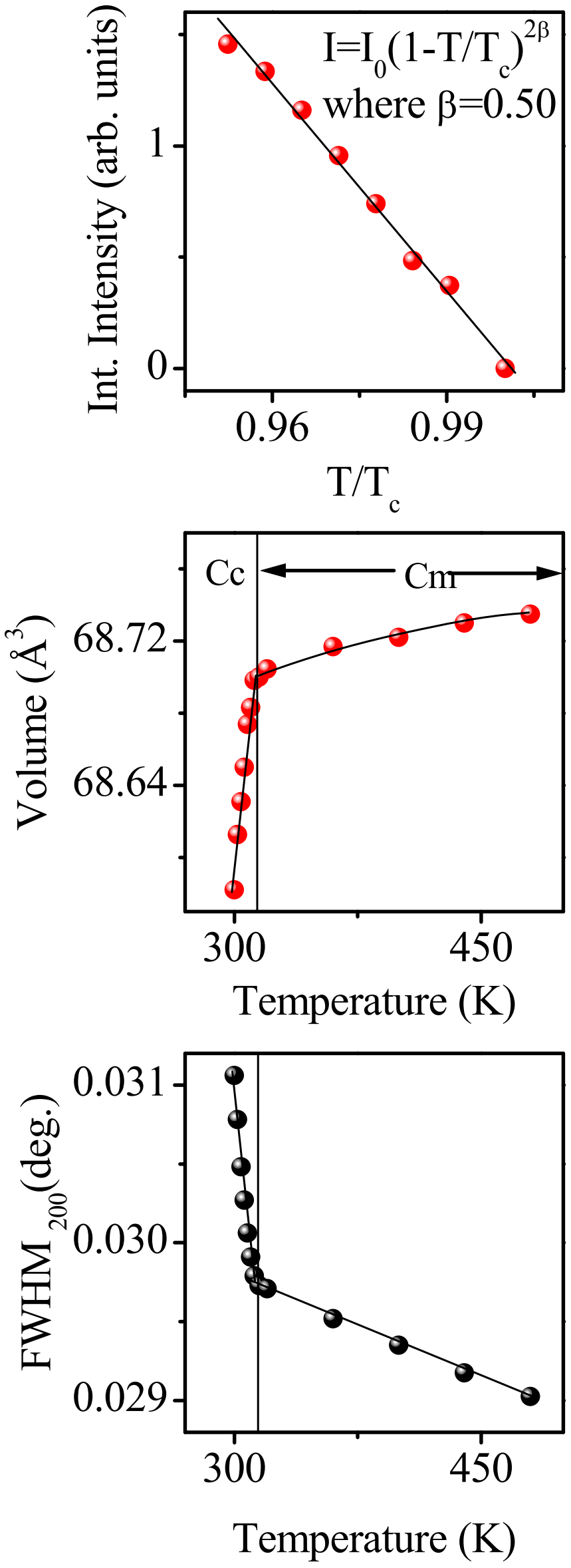

Fig. 4

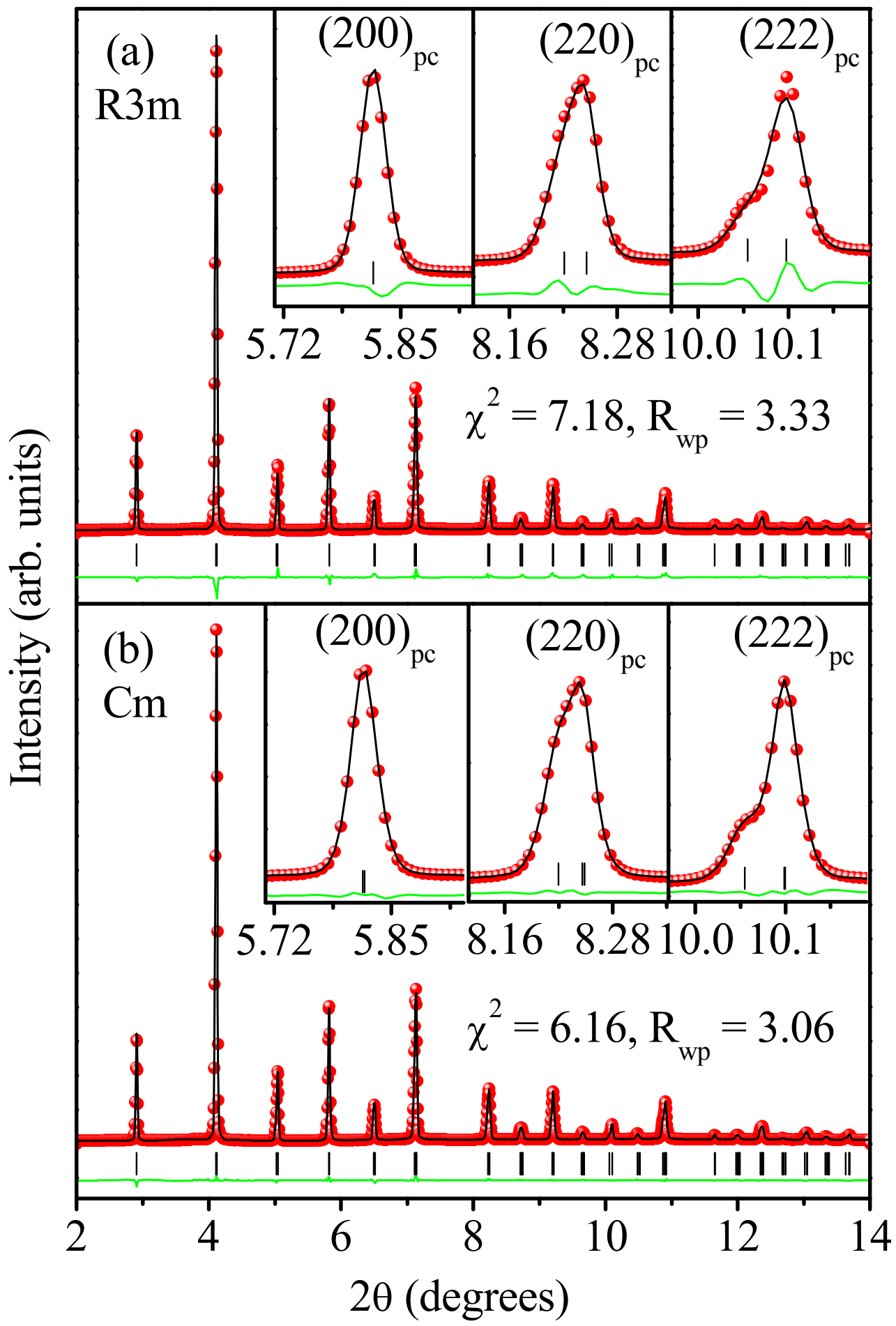

Fig. 5

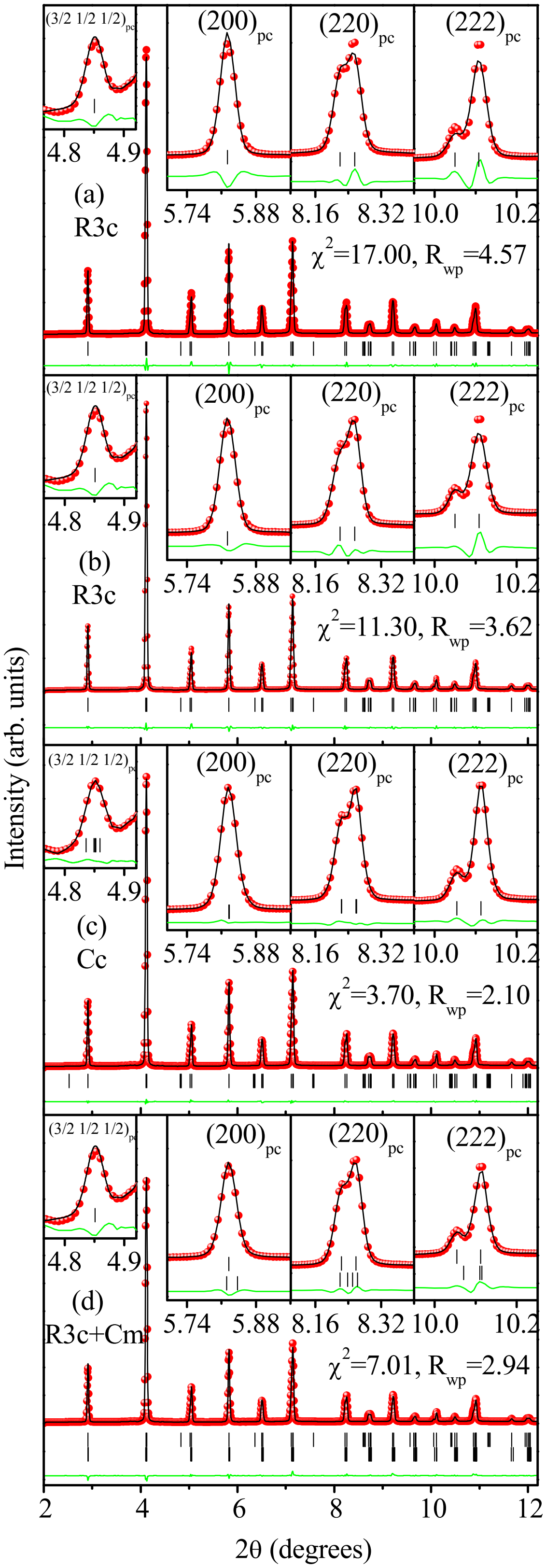

Fig. 6

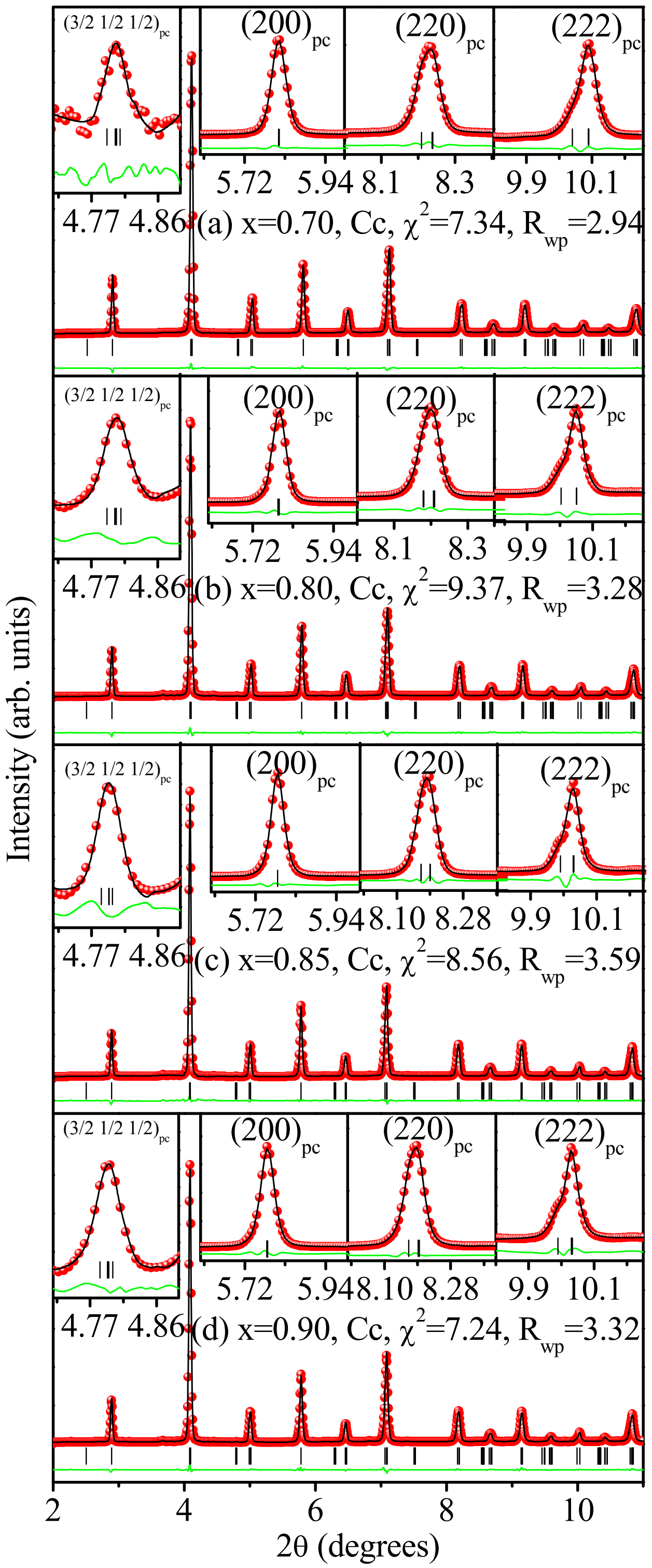
(3/2 1/2 1/2)pc
(200)pc
(220)pc
(222)pc
(a) x=0.70, Cc, χ2=7.34, Rwp=2.94
(b) x=0.80, Cc, χ2=9.37, Rwp=3.28
(c) x=0.85, Cc, χ2=8.56, Rwp=3.59
(d) x=0.90, Cc, χ2=7.24, Rwp=3.32
Intensity (arb. units)
2θ (degrees)

Fig. 7

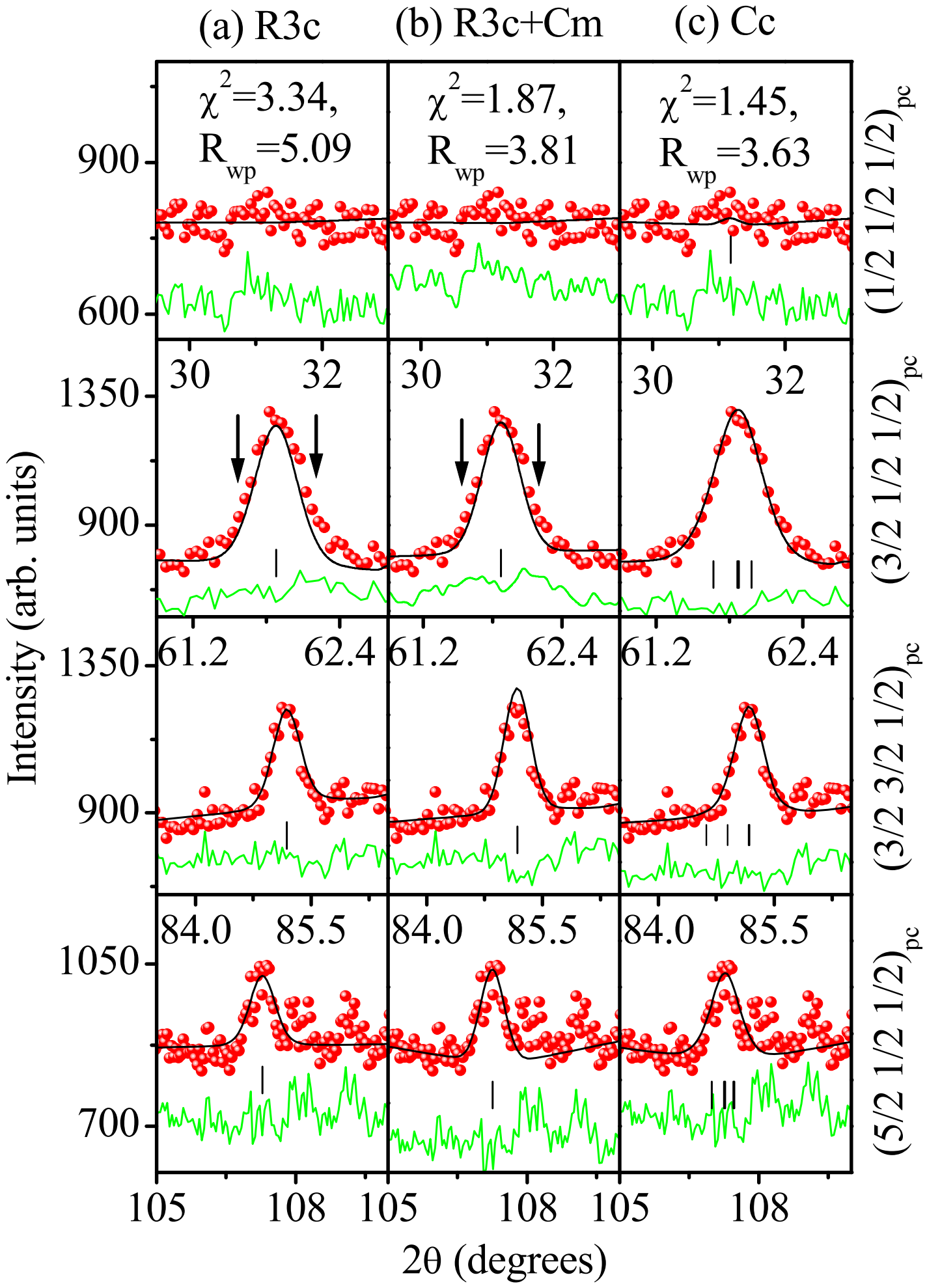

Fig. 8

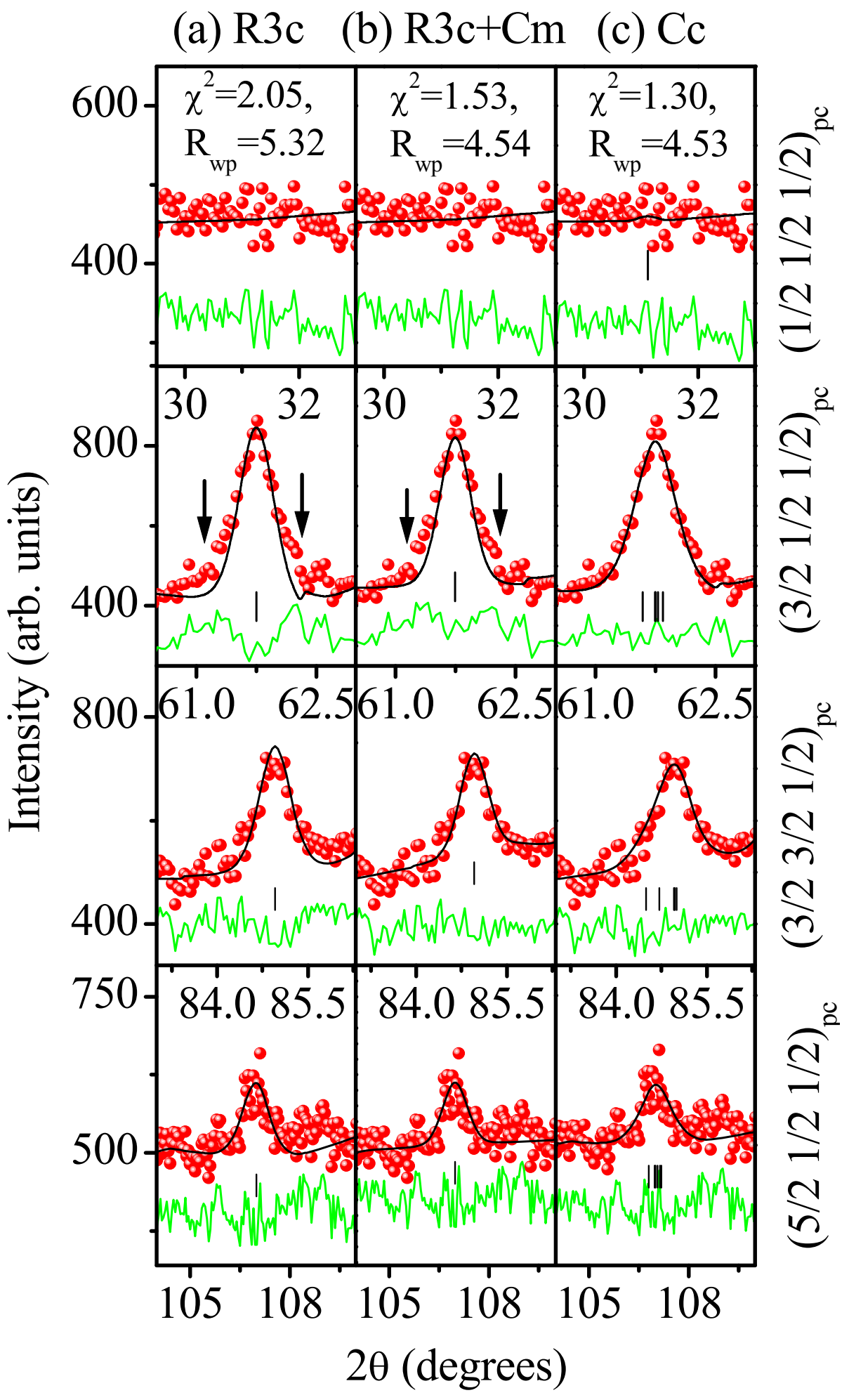

Fig. 9

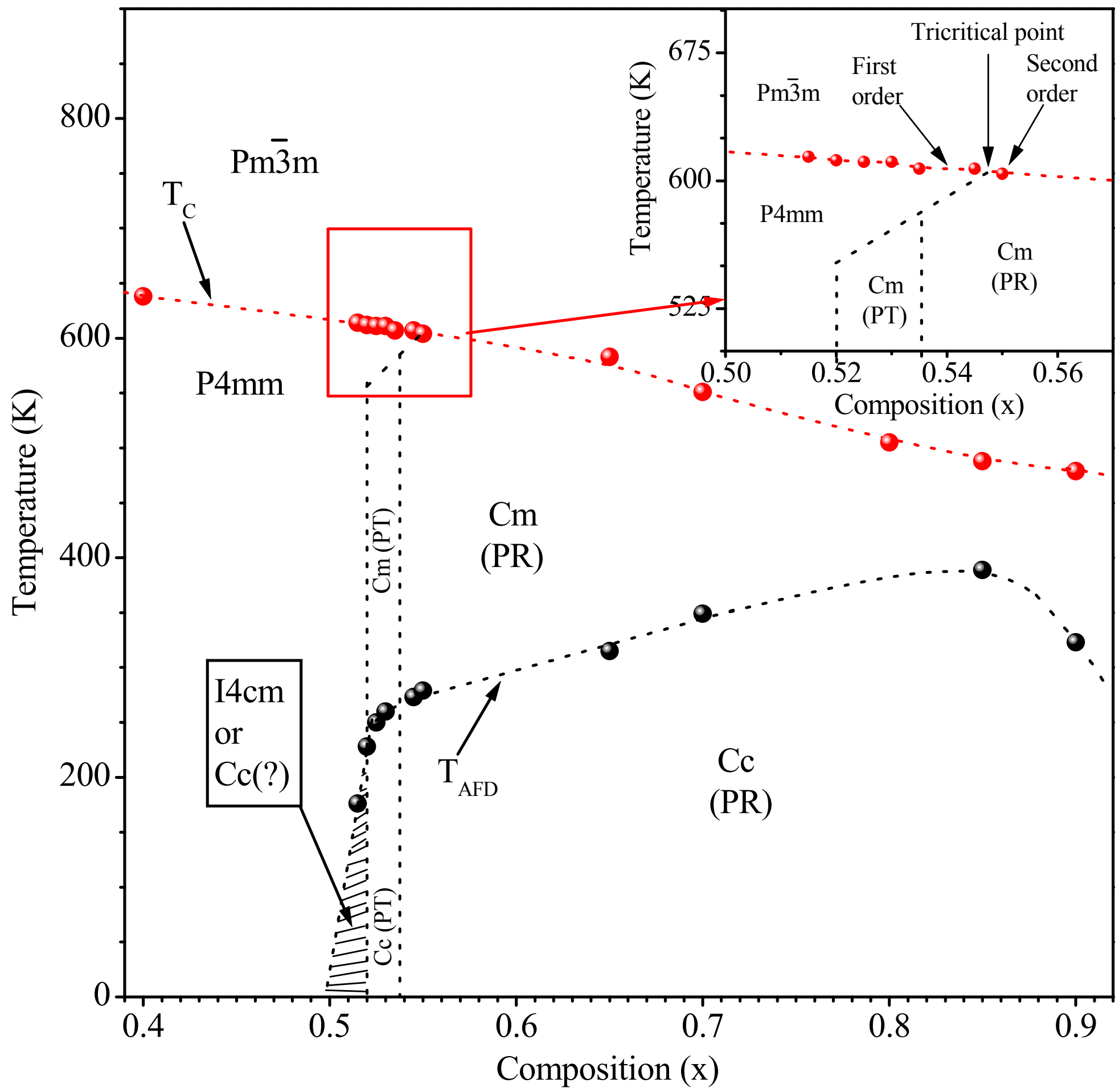

Fig. 10